\documentstyle[12pt]{article}
\begin{document}
\newcommand{\beq}{\begin{equation}}
\newcommand{\eeq}{\end{equation}}
\newcommand{\bea}{\begin{eqnarray}}
\newcommand{\eea}{\end{eqnarray}}
\newcommand{\half}{\frac{1}{2}}
\newcommand{\quart}{\frac{1}{4}}
\newcommand{\ihalf}{\frac{i}{2}}
\newcommand{\ehalf}{\frac{e}{2}}
\newcommand{\khalf}{\frac{\kappa}{2}}
\newcommand{\ikhalf}{\frac{i\kappa}{2}}
\newcommand{\metric}{\sqrt{-g}}
\newcommand{\G}{\Gamma}
\newcommand{\aslash}{\not\!\! A}
\newcommand{\bslash}{\kern-.01em\hbox{\raise.25ex\hbox{$/$}\kern-.52em$b$}}
\newcommand{\cslash}{\kern-.01em\hbox{\raise.25ex\hbox{$/$}\kern-.52em$c$}}
\newcommand{\sslash}{/\kern-.52em s}
\newcommand{\kslash}{\not\! k}
\newcommand{\dslash}{\not\!\partial}
\newcommand{\ddslash}{\not\!\! D}
\newcommand{\derboth}{\stackrel{\leftrightarrow}{\partial}}
\newcommand{\psinapsi}{\bar{\psi}_{\rho}\hat{\nabla}_{\mu}\psi_{\sigma}}
\newcommand{\dmu}{\partial_{\mu}}
\newcommand{\dnu}{\partial_{\nu}}
\renewcommand{\a}{\alpha}
\renewcommand{\b}{\beta}
\newcommand{\g}{\gamma}
\newcommand{\e}{\epsilon}
\renewcommand{\l}{\lambda}
\newcommand{\oz}{\bar{z}}
\newcommand{\disp}{\displaystyle}
\newcommand{\ze}{\oz}
\newcommand{\cd}{\cal D}
\newcommand{\opsi}{\bar\Psi}
\newcommand{\intppll}{\int{\cal D}\bar\Psi{\cal D}\Psi{\cal 
D}\bar\l{\cal D}\l}
\newcommand{\intpp}{\int{\cal D}\bar\Psi{\cal D}\Psi}
\newcommand{\intll}{\int{\cal D}\bar\l{\cal D}\l}
\newcommand{\intppllb}{\int{\cal D}\bar\Psi{\cal D}\Psi{\cal 
D}\bar\l{\cal D}\l{\cal D}b_{\mu}}
\newcommand{\intppb}{\int{\cal D}\bar\Psi{\cal D}\Psi{\cal D}b_{\mu}}
\newcommand{\intppc}{\int{\cal D}\bar\Psi{\cal D}\Psi{\cal D}c_{\mu}}
\newcommand{\intlla}{\int{\cal D}\bar\l{\cal D}\l{\cal D}A_{\mu}}
\newcommand{\intllab}{\int{\cal D}\bar\l{\cal D}\l{\cal 
D}A_{\mu}{\cal D}b_{\mu}}
\newcommand{\intppabc}{\int{\cal D}\bar\Psi{\cal D}\Psi{\cal 
D}A_{\mu}{\cal D}b_{\mu}{\cal D}c_{\mu}}
\newcommand{\intab}{\int{\cal D}A_{\mu}{\cal D}b_{\mu}}
\newcommand{\inta}{\int{\cal D}A_{\mu}}
\newcommand{\intb}{\int{\cal D}b_{\mu}}
\newcommand{\intbf}{\int{\cal D}b_{\mu}{\cal D}\phi}
\newcommand{\intef}{\int{\cal D}\eta{\cal D}\phi}
\newcommand{\intf}{\int{\cal D}\phi}
\newcommand{\lpp}{\left.\left.}
\newcommand{\rppnn}{\right.\right.\nonumber\\}
\newcommand{\opar}{\bar\epsilon}
\newcommand{\tres}{\;\;\;}
\hyphenation{bo-so-ni-za-tion}

\title{Supersymmetry and Bosonization in three dimensions}
\author{Jos\'e D. Edelstein\thanks{E-mail address: 
edels@venus.fisica.unlp.edu.ar}~ and Carlos N\'u\~nez\thanks{E-mail 
address: nunez@venus.fisica.unlp.edu.ar} \\
Departamento de F\'{\i}sica, Universidad Nacional de 
La Plata \\ C.C. 67, (1900) La Plata, Argentina}

\maketitle
\setcounter{page}{1}

\begin{abstract}
We discuss on the possible existence of a supersymmetric invariance
in purely fermionic planar systems and its relation to
the fermion-boson mapping in three-dimensional quantum field
theory.
We consider, as a very simple example, the bosonization of free 
massive fermions and show that, under certain conditions on the 
masses, this model displays a supersymmetric-like invariance in the 
low energy regime.
We construct the purely fermionic expression for the supercurrent 
and the non-linear supersymmetry transformation laws.
We argue that the supersymmetry is absent in the limit of massless
fermions where the bosonized theory is non-local.
\end{abstract}

It is a well-known fact that two-dimensional models admit non-linear
realizations of the supersymmetry algebra that involve only fermions
\cite{PvN}. Indeed, in two dimensions one can hope to represent
the full supersymmetry algebra solely in terms of either bosonic
or fermionic fields by means of bosonization (fermionization)
\cite{SC,SM}. This property of two-dimensional purely fermionic 
models was originally analyzed by Witten \cite{EW} and then
investigated in detail by Aratyn and Damgaard \cite{HAPHD}.
Several interesting fermionic models were then investigated
and shown to provide a non-linear realization of 
supersymmetry \cite{PdVVGKJLPPR,FASMLT}. 

In order to extend this result to higher dimensional models, one should
be provided with a bosonization recipe valid for $d>2$. It was not until
very recently that a fermion-boson mapping has been established in $d=3$
dimensions \cite{CPBFQ,CPBCALFQ,EFFAS,FAS}. 
The resulting bosonization rules for
fermionic currents, either for free or interacting models, result to 
be the natural extension of the two-dimensional ones \cite{EFFAS,FAS}:
\beq
j_{\mu} = -i\bar\Psi\gamma_{\mu}\Psi \rightarrow
\frac{1}{\sqrt{4\pi}}\e_{\mu\nu\a}\partial_{\nu}A_{\a} =
\frac{1}{\sqrt{4\pi}}{\cal H}_{\mu} ~,
\label{curr}
\eeq
where $\Psi$ is a Dirac fermion, $A_{\a}$ is a vector field and 
${\cal H}_{\mu}$ is the corresponding dual field strength. 
It is understood that we are
working in three-dimensional euclidean space-time. The exact bosonic
Lagrangian equivalent to a given fermionic one, however, cannot be
obtained unless some approximations are taken into account.

In the present work we plan to address the problem of bosonization
(fermionization) of three-dimensional supersymmetric models. 
The study of supersymmetric planar systems has recently
shown to be of interest in the domain of cosmology \cite{see}.
It could also be relevant to the supermembrane theory, whose world 
volume action should be a three-dimensional supersymmetric
field theory.

We consider in this letter, as a very simple toy
model, the case of free massive fermions, as it displays all the 
interesting novel features of our construction: 
a non-linear realization of supersymmetry in terms of fermionic 
variables at low energy, which is broken in the limit
of massless fermions.
We will report separately on the uses of our technique to treat 
more involved interacting systems \cite{JDECN}.

We start from a purely fermionic Abelian model
consisting of two non-interacting massive Dirac and Majorana
fermions. 
We carry out the bosonization of the Dirac fermion
using path-integral methods, arriving to an effective theory of a
Majorana spinor and a vector field whose kinetic action consists
of a couple of highly involved functions of the Laplacian. 
We show that the resulting lagrangian displays a supersymmetric
invariance in the low energy regime, provided the original 
fermionic masses satisfy a definite ratio.
We compute the purely fermionic supersymmetry transformation 
laws taking advantage of the exact bosonization recipe for
currents. 
When massless fermions are considered, instead, the resulting
bosonized model is not in principle supersymmetric. 
Finally, we add external sources to our system
and find a very simple relation between correlation functions
of currents computed in the purely fermionic system and those
calculated in the corresponding supersymmetric model.

~

Let us begin by considering a simple purely fermionic model 
in three-dimensional Euclidean
space-time described by the following Lagrangian density:
\beq
{\cal L}_{f} = \half\bar\Psi(\dslash + m_{\Psi})\Psi +
\half\bar\l(\dslash + m_{\l})\l ~,
\label{laguno}
\eeq
where $\Psi$ is a massive Dirac spinor and $\l$ is a massive
Majorana fermion. The corresponding partition function 
\beq
Z_{f} = \intppll\exp\left[-\half\int d^3x \left( 
\bar{\Psi}(\dslash + m_{\Psi})\Psi + \bar\l(\dslash + 
m_{\l})\l \right) \right] ~,  
\label{zfer1}
\eeq
can be trivially factored as $Z_{f} = Z_{\Psi} Z_{\l}$, with
\beq
Z_{\Psi} = \intpp\exp\left[-\half\int d^3x \bar{\Psi}(\dslash +
m_{\Psi})\Psi\right] ~,  
\label{zpsi}
\eeq
\beq
Z_{\l} = \intll\exp\left[-\half\int d^3x \bar\l(\dslash +
m_{\l})\l\right] ~.
\label{zl}
\eeq

Let us perform the change on Dirac fermionic variables $\Psi = 
g(x) \Psi'$ and $\bar\Psi = \bar\Psi' g^{-1}(x)$, with $g(x) = 
\exp[i\theta(x)]$ an element of U(1). One can always define for 
Dirac fermions a path-integral measure invariant under such
transformations. 
After the change of variables, the partition function 
$Z_{\Psi}$ becomes
\beq
Z_{\Psi} = \intpp\exp\left[-\half\int d^3x \bar{\Psi}(\dslash 
+ i\!\dslash\theta + m_{\Psi})\Psi\right] 
\label{zfer2}
\eeq
(we have omitted primes in the new fermionic variables). 
In this formula, $\dmu\theta$ can be simply seen as a flat connection.
Thus, it can be replaced by a gauge field connection $b_\mu$ provided 
a flatness condition is imposed,
\beq
b_\mu = \dmu\theta \tres \mbox{with} \tres
\bar{\cal F}_{\mu} = \e_{\mu\nu\gamma}\partial_{\nu}b_{\gamma} = 0 ~.
\label{gauge}
\eeq
Now, following the treatment of Ref.\cite{FAS}, the partition function 
(\ref{zfer2}) becomes
\[ Z_{\Psi} = \intab det(\ddslash[b] + 
m_{\Psi})\exp\left[i\!\int 
d^3\! x A_{\mu}\bar{\cal F}_{\mu}\right] ~, \]
where $D_{\mu}[b] = \partial_{\mu} + ib_{\mu}$ is the covariant
derivative. It is the Lagrange multiplier $A_{\mu}$ that would
become the bosonic equivalent of the original Dirac fermion $\Psi$. 
Of course, one has still to perform the $b_\mu$ integration
to obtain explicitely the dynamics of this vector field.
To this end, we should first compute the fermion determinant in
three dimensions as exactly as we can.
In the limit of very massive fermions, it is a well-known result
that the fermion determinant leads to an effective action given
by a Chern-Simons term \cite{ANRRDPSR,SDRJST}.
More recently, this calculation has been done for any value of the
fermion mass by making a quadratic expansion in 
powers of the $b_{\mu}$-field, with the following result 
\cite{KDR,IJRACDFJAZ,DGBCDFLEO}:
\beq
-\log\det(\ddslash[b] + m_{\Psi}) \cong \frac{1}{4}\int d^3x
\left(b_{\mu\nu}F(-\partial^2)b_{\mu\nu} + 
2ib_{\mu}G(-\partial^2)\bar{\cal F}_{\mu} \right) ~,
\label{result}
\eeq
where $b_{\mu\nu} = \dmu b_{\nu} - \dnu b_{\mu}$. 
Both contributions in Eq.(\ref{result}) can be traced to 
come from the parity-conserving and parity-violating pieces of the 
vacuum-polarization tensor \cite{SDRJST}. Functions 
$F(-\partial^2)$ and $G(-\partial^2)$ can be given through their 
momentum-space representations ${\tilde F}$ and ${\tilde G}$ as 
follows \cite{KDR,IJRACDFJAZ,DGBCDFLEO,JCLGCNFAS}:
\beq
\frac{4\pi k^2}{|m_{\Psi}|}{\tilde F} (k) = 
1 - \disp{\frac{1 \,-\, \disp{\frac{k^2}{4 
m_{\Psi}^2}}}{\left(\disp{\frac{k^2}{4 
m_{\Psi}^2}}\right)^{1/2}}} 
\, \arcsin\left(1\,+ \, \frac{4 m_{\Psi}^2}{k^2}\right)^{-1/2} 
\label{efek}
\eeq
which results to be regularization independent, and
\beq
{\tilde G} (k) \;=\; \frac{q}{4 \pi} \,+\, 
\frac{m_{\Psi}}{2 \pi \mid k \mid} \, \arcsin\left(1 
\, + \, \frac{4 m_{\Psi}^2}{k^2}\right)^{-1/2} ~,
\label{gek}
\eeq
which is regularization dependent. Indeed, the parameter
$q$ can assume any integer value \cite{JFTK,REGSGLRFAS}.

~

The quadratic expansion of the fermion determinant must be taken
with care. There is no small parameter in the theory that allows
a sensible perturbative expansion of the path-integral. Therefore,
generically, it has no physical sense to cut the whole series of
$b_{\mu}$-insertions into the fermion loop leading to the
value of the determinant. In spite of this fact, it is important
to mention that there are some cases in which this difficulty
can be overcomed, one of them being the limit of very heavy
fermions. If one is to analyse the low energy regime of the 
theory, imposed by letting $m_{\Psi}\to\infty$, one can take
profit of the well-known fact \cite{CPBCALFQ} that the series
expansion of the fermion determinant can be accomodated as an
expansion in powers of the inverse fermion mass. The truncation
of the late series is well-defined in the low-energy limit. Indeed, 
it leads to a local effective lagrangian dominated by those terms 
that have the lowest scaling dimension \cite{CPBCALFQ}, which are
precisely those appearing in (\ref{result}), when evaluated in 
the appropriate limit.

In order to illustrate this argument, one can compute the leading
(parity conserving) contribution in $m_{\Psi}^{-1}$ of the fermion 
loop with four $b_{\mu}$-insertions, which result
to be proportional to $m_{\Psi}^{-5}(F_{\mu\nu}^2)^2$. This result 
can be also obtained following the derivative expansion approach 
\cite{IJRACMFBLHDJOC}, as was recently analysed in \cite{IJRACDFFDM}.
Taking into account these remarks, one must keep in mind in what
follows that sensible results are those emerging in the low
energy limit. It is worthwhile to mention, however, that there
is another situation in which the quadratic approximation of the
fermion determinant is physically relevant for any value of the
mass (including the limit $m_{\Psi} \to 0$). This is the calculation
of the current algebra first done in Ref.\cite{JCLGCNFAS}, for
which the following terms in the expansion are irrelevant. In 
fact, the quadratic approximation was explicitely shown to be
enough to reproduce the result of the current commutators that
corresponds to that of the dual fermion theory. We will come back 
to this sort of discussion below.

~

Using the quadratic approximation to compute the fermionic 
determinant (\ref{result}), it is possible to perform the exact 
integration of the $b_{\mu}$ field \cite{DGBCDFLEO}, the resulting
partition function being:
\beq
Z_{f} = \inta\exp\left[-\!\int d^3x
\left(\frac{1}{4}F_{\mu\nu}C_{1}(-\partial^2)F_{\mu\nu} - \ihalf 
A_{\mu}C_{2}(-\partial^2){\cal H}_{\mu} \right) \right] Z_{\l} ~,
\label{zfer5}
\eeq
with $F_{\mu\nu} = \partial_{\mu} A_{\nu} - \partial_{\nu} 
A_{\mu}$.
Concerning $C_1$ and $C_2$, they are given through their 
momentum-space representations ${\tilde C_1}$ and ${\tilde C_2}$ 
as follows,
\beq
{\tilde C_1} (k) = \frac{|u|^2 {\tilde F}}{k^{2} {\tilde F}^2 + 
{\tilde G}^2} \;\;\; , \;\;\; {\tilde C_2} (k) = 
\frac{|u|^2 {\tilde G}}{k^{2} {\tilde F}^2 + {\tilde G}^2} ~,
\label{ces}
\eeq
where $u$ is an arbitrary function of the momentum, arising in the 
diagonalization procedure. Current correlation functions are 
$u$-independent \cite{JCLGCNFAS}. 
Had we added in the partition function (\ref{zfer1}) an external 
source $s_{\mu}$ covariantly coupled to the fermionic current
\beq
Z_{f}[s] = \intppll\exp\left[-\half\int d^3x \left( 
\bar{\Psi}(\ddslash[s]  + m_{\Psi})\Psi + \bar\l(\dslash + 
m_{\l})\l \right) \right] ~,  
\label{wcur}
\eeq
we would obtain --after the same procedure is applied--, the 
bosonized expression:
\bea
Z_{f}[s] & = & \inta\exp\left[-\!\int d^3x
\left(\frac{1}{4}F_{\mu\nu}C_{1}(-\partial^2)F_{\mu\nu} 
- \ihalf A_{\mu}C_{2}(-\partial^2){\cal H}_{\mu} \rppnn
& + & \lpp iu s_{\mu}{\cal H}_{\mu} \right)\right] Z_{\l} ~.
\label{concorr}
\eea
Note that, choosing the value $u = i/4\sqrt{\pi}$,
the preceding equation makes evident the exactness of
the bosonization recipe for currents (\ref{curr}) in correspondence
with our previous discussions \cite{JCLGCNFAS}. 

~

If we let the mass $m_{\Psi}$ go to infinity,
${\tilde F}(k)$ and
${\tilde G}(k)$ take simpler expressions such that the 
leading order contributions to ${\tilde C_1}$ and ${\tilde C_2}$ are
\beq
{\tilde C_1} (k) \rightarrow \frac{1}{3\mid m_{\Psi} \mid} \tres
\mbox{and} \tres {\tilde C_2} (k) \rightarrow -1 ~,
\label{cesinf}
\eeq
and the partition function then becomes
\bea
Z_{m_{\Psi}\to\infty} & = & \intlla\exp\left[-\int d^3x
\left(\frac{1}{12|m_{\Psi}|}F_{\mu\nu}^2 + \frac{i}{2} 
A_{\mu}{\cal H}_{\mu} \rppnn
& + & \lpp \half\bar\l(\dslash + 
m_{\l})\l \right) \right] ~.
\label{zlow1}
\eea
After a rescaling of the bosonized field $A_{\mu}$,
\beq
A_{\mu} = \sqrt{3m_{\Psi}}A_{\mu}^{\prime} ~,
\label{scale}
\eeq
the partition function (\ref{zlow1}) simply reads
\bea
Z_{m_{\Psi}\to\infty} & = & \intlla\exp\left[-\int d^3x
\left(\frac{1}{4}F_{\mu\nu}^{\prime}F_{\mu\nu}^{\prime} + 
\frac{3im_{\Psi}}{4}\e_{\mu\nu\g}F_{\mu\nu}^{\prime}A_{\g}^{\prime} 
\rppnn
& + & \lpp \half\bar\l(\dslash + m_{\l})\l
\right) \right] ~,
\label{zlow2}
\eea
whose bosonic sector is given by the Maxwell-Chern-Simons theory 
\cite{SDRJST}. The partition function (\ref{zlow2}) describes
a free vector boson with a gauge invariant mass which is three times
bigger than the mass of the Dirac fermion that was at its origin,
$m_{bos} = 3m_{\Psi}$, and a free massive Majorana fermion with
$m_{fer} = m_{\l}$. It would be possible to place these fields together
into the same representation of $N=1$ supersymmetry provided their mass
is equal. In fact, choosing 
\beq
m_{\l} = 3m_{\Psi} ~,
\label{cond}
\eeq
the partition function (\ref{zlow2})
is invariant under the supersymmetry transformations
\beq
\delta A_{\mu}^{\prime} = \opar\g_{\mu}\l \tres , \tres \delta 
\l = \ihalf\e_{\mu\nu\g}\g_{\mu}F_{\nu\g}^{\prime}\e ~,
\label{susys}
\eeq
with $\e$ a real spinor parameter.
As this is a free theory the invariance of the action does not, by
itself, establish supersymmetry. It is easy to 
check that the commutator of two transformations (\ref{susys})
produces a traslation on the fields. Unfortunately this does not say 
much about the possibility of obtaining the same kind of results
once interactions are included: determining non-trivial aspects of 
this Supersymmetry certainly requires thorough investigations on the 
ground of interacting theories. In this respect, it is worth to 
mention that some very simple 
interacting systems can be handled whithin this approach 
arriving to analogous results \cite{JDECN}. 

~

We have shown, using the path-integral bosonization framework, the 
equivalence at low energy between the purely fermionic model and the 
Max\-well-Chern-Simons supersymmetric one, provided the fermion masses
satisfy a definite ratio. To our knowledge, this kind of equivalence
has not been studied before in planar systems neither in higher
dimensional models.
It is worthwhile to note 
that this limiting case could also be considered when an external
source is coupled to the fermionic current. Consequently, from
eq.(\ref{wcur}), the correlation function of these currents in 
the low energy limit, after bosonization, are
simply given by
\beq
<j_{\mu}(x)j_{\nu}(y)> = \frac{e^2}{4\pi}<{\cal H}_{\mu}(x){\cal 
H}_{\nu}(y)>_{SUSY} ~,
\label{correl}
\eeq
where the subscript $SUSY$ means that the correlation function at
the r.h.s. of (\ref{correl}) is computed in the supersymmetric
theory with the partition function given in (\ref{zlow2}).
Thus, the purely fermionic free correlator can be calculated as 
the correlation function of the dual electromagnetic field strength 
in the $N=1$ supersymmetric Maxwell-Chern-Simons theory.

~

In order to further study the supersymmetry 
of the purely fermionic model, let 
us compute the Noether supercurrent corresponding to the Lagrangian in
eq.(\ref{zlow2}). It is given by the following expression
\beq
{\cal J}_{\mu} = i{\cal H}_{\nu}^{\prime}\g_{\nu}\g_{\mu}\l ~,
\label{supcurr1}
\eeq
whose vacuum expectation value can be written by adding sources as
\beq
<{\cal J}_{\mu}^{\g}> = 
-i\left[\frac{16\pi}{3m_{\Psi}}\right]^{1/2}
(\g_{\nu}\g_{\mu})^{\g}_{~~\b}
\frac{\delta^2\log{Z}[\eta,\bar\eta,s]}{\delta 
s_{\nu}\delta\bar\eta_{\b}}\vert_{\eta,\bar\eta,s=0} ~,
\label{mean}
\eeq
where $Z[\eta,\bar\eta,s]$ is the partition function (\ref{zlow2})
coupled to external sources given by
\bea
Z[\eta,\bar\eta,s] & = & \intlla\exp\left[-\int d^3x \left( 
\frac{1}{4}F_{\mu\nu}^{\prime}F_{\mu\nu}^{\prime} +
\frac{3im_{\Psi}}{4}\e_{\mu\nu\g}F_{\mu\nu}^{\prime}A_{\g}^{\prime} 
\rppnn & - & \lpp 
\left(\frac{3m_{\Psi}}{16\pi}\right)^{1/2}\!s_{\mu}{\cal 
H}_{\mu}^{\prime} + \half\bar\l(\dslash + m_{\l})\l + \bar\eta\l + 
\bar\l\eta \right) \right] ~.
\label{mean1}
\eea
We can fermionize back the low energy action (\ref{mean1}),
following the procedure we have just discussed, to end with
\bea
Z[\eta,\bar\eta,s] & = & \intppll\exp\left[-\half\int d^3x \left( 
\bar{\Psi}(\ddslash[s] + m_{\Psi})\Psi \rppnn & + & \lpp
\bar\l(\dslash + m_{\l})\l + \bar\eta\l + \bar\l\eta \right) 
\right] ~,
\label{mean3}
\eea
this leading to the purely fermionic expression for the vacuum
expectation value of the supercurrent
\beq
<{\cal J}_{\mu}^{\g}> = 
\a^{-1/2}<\bar\Psi\g_{\nu}\Psi(\g_{\nu}\g_{\mu}\l)^{\g}> ~,
\label{mean2}
\eeq
with $\a = 3m_{\Psi}/4\pi$.
The expression (\ref{mean2}) gives, in the low energy limit, the 
vacuum expectation value of the supercurrent that corresponds to the 
supersymmetric invariance of the purely fermionic Lagrangian 
(\ref{laguno}) subjected to the condition (\ref{cond}). 

~

If eq. (\ref{mean2}) were valid at the operator level
we would easily be able to show that it is a divergenceless current. However,
being the current a composite operator, one must be very careful with 
potential Schwinger-like terms coming from the time ordered products.
Using the previous expression for the current, and its $\mu =0$ 
component, the supercharge, one can obtain the following fermionic 
infinitesimal transformation laws:
\beq
\delta_{\e}\l = \a^{-1/2}\bar\Psi\g_{\nu}\Psi\g_{\nu}\e ~,
\label{traf1}
\eeq
\beq
\delta_{\e}\Psi = \a^{-1/2}(\bar\l\e \Psi + \bar\l\g_{\mu}\e 
\g_{\mu}\Psi) ~. 
\label{traf2}
\eeq
Transformations (\ref{traf1}) and (\ref{traf2}) provide a 
non-linear realization of the supersymmetric invariance that is
present in the low energy regime of the purely fermionic model
given in eq.(\ref{laguno}). 
It is inmediate to see that the rule (\ref{traf1}), 
is nothing but the straight fermionization of the photino 
transformation law given in Eq.(\ref{susys}). 
Concerning Eq.(\ref{traf2}), it is obtained by acting with the 
supercharge constructed from (\ref{supcurr1}) over the Dirac 
spinor $\Psi$, not as in $1+1$-dimensional models where it is 
interestingly computed from the chiral anomaly \cite{HAPHD} (a 
procedure which is, of course, impossible in our case).

~

The `duality' between the purely fermionic model (\ref{laguno})
and the $N=1$ supersymmetric Maxwell-Chern-Simons one (\ref{zlow2})
--after condition (\ref{cond}) is imposed--, was established in the
path-integral framework by considering the low energy limit of the
fermionic determinant. 
It is interesting to explore whether this kind of correspondence 
is still present in the high energy regime (thought of as the limit
of massless fermions, $m_{\Psi} \to 0$). 
As we extensively discussed before, this limit is somehow problematic
in the sense that the quadratic approximation to the fermionic 
determinant could have no sense. However, we will briefly analyse
this case, taking into account that the resulting bosonized action
of massless fermions obtained by this method --in the quadratic
approximation-- was shown to give a sensible physical answer \cite{ECM}.
Also, the quadratic approximation that we followed has been shown to
be enough in order to reproduce the current algebra of the fermionic
system in terms of bosonic variables \cite{JCLGCNFAS}.
Let us then consider in what follows the 
case where $m_{\Psi}$ is almost vanishing. In this case we have the 
following behaviour for ${\tilde F}$ and ${\tilde G}$:
\beq
{\tilde F} (k) \rightarrow \frac{1}{16\mid k \mid} \tres \mbox{and}
\tres {\tilde G} (k) \rightarrow \frac{q}{4\pi} \equiv
\frac{1}{16} \cot\varphi ~.
\label{efege0}
\eeq

Inserting these expressions into Eqs.(\ref{zfer5}) and (\ref{ces}), 
we are led to:
\bea
Z_{m_{\Psi}\to 0} & = & 
\inta\exp\!\left[-\frac{\sin^2\!\varphi}{\pi}\int\! 
\left(F_{\mu\nu}\frac{1}{\sqrt{-\partial^2}}F_{\mu\nu} \rppnn
& - & \lpp 2i\cot\varphi\,\e_{\mu\nu\g}A_{\mu}F_{\nu\g} 
\right) d^3x \right] Z_{\l} ~,
\label{zhigh1}
\eea
a non-local theory which is by no means supersymmetric.
The fermionized supersymmetric invariance of the free massive
fermion system seems to be lost at high energies.
It is worth to comment at this point that non-local terms as that
of Eq.(\ref{zhigh1}) can be traced up as coming from a 
four-dimensional model, after a dimensional reduction procedure
is applied \cite{ECM,JDEJJGCNFAS}.

~

Let us end this letter by stressing its main results. We have 
presented a very simple purely fermionic model and showed that it 
is supersymmetric in the low energy regime, provided the fermion
masses satisfy a definite ratio.
We were able to construct the non-linear supersymmetry 
transformations corresponding to this invariance in terms of
purely fermionic variables. 
In the limit of massless fermions, instead, the supersymmetry
seems to be absent and the bosonized theory results to be non-local.
It would be of interest to study the way in which the transition
between both phases of the theory occurs.

~

In order to clarify our discussion we have considered in this letter 
a free fermionic model. 
It is however possible to add certain interaction terms without
spoiling the basic conclusions to which we have arrived 
here \cite{JDECN}.
The possible generalization of these results to 
higher-dimensional systems deserve further 
investigations.
We hope to report on these issues elsewhere.

~

We would like to thank Adri\'an Lugo, Carlos Na\'on
and Fernando Quevedo for helpful comments.
We are also pleased to thank the referee for most valuable
remarks that significantly improved the presentation.
This work was partially supported by CONICET, Argentina.

\end{document}